\documentclass[a4paper]{JHEP}

\usepackage{epsfig}

\newcommand{\aj}[3]{\emph{AJ}\ {\bf #1} (#2) #3}
\newcommand{\apj}[3]{\emph{ApJ}\ {\bf #1} (#2) #3}
\newcommand{\nat}[3]{\emph{Nature}\ {\bf #1} (#2) #3}
\newcommand{\mnras}[3]{\emph{MNRAS}\ {\bf #1} (#2) #3}
\newcommand{\astpar}[3]{\emph{Astropart. Phys.}\ {\bf #1} (#2) #3}
\newcommand{\na}[3]{\emph{New A}\ {\bf #1} (#2) #3}

\def\eps{{\cal E}}

\title{Velocity distributions and annual-modulation 
signatures of weakly-interacting massive particles}

\author{Piero Ullio and Marc Kamionkowski\\
	Mail Code 130-33, California Institute of Technology,
	Pasadena, CA 91125 \\
	E-mail:  \email{piero@tapir.caltech.edu},
        \email{kamion@tapir.caltech.edu}}

\abstract{
An annual modulation in the event rate of the NaI detector of the
{\sc Dama} collaboration
has been used to infer the existence of particle dark matter in
the Galactic halo. Bounds on the WIMP mass and WIMP-nucleon cross 
section have been derived. These analyses have assumed
that the local dark-matter velocity distribution is either isotropic
or has some bulk rotation.
Here we consider the effects of possible structure in the WIMP 
velocity distribution on the annual-modulation amplitude.
We show that if we allow for a locally anisotropic velocity 
dispersion tensor, the interpretation of direct detection 
experiments could be altered significantly.
We also show that uncertainties in the velocity distribution
function that arise from uncertainties in the radial density
profile are less important if the velocity dispersion is assumed 
to be isotropic.}

\preprint{}


\begin{document}

\section{Introduction} 

Weakly interacting massive particles (WIMPs) are among the leading 
candidates for dark matter in galactic halos.
Such particles arise naturally in extensions to the standard model
(SM) of particle physics; an example is the neutralino, plausibly
the lightest superpartner in supersymmetric versions of the SM.
Massive particles whose coupling with lighter
SM particles have interactions of electroweak strength have a
cosmological abundance of order the critical density of the Universe.
Hence, WIMPs appear naturally as dark-matter candidates. 
The possibility to link these two apparently separate problems
(electroweak symmetry breaking and dark matter)
was realized a couple of decades ago, and since then the search for 
WIMPs in the Milky Way halo has been a major endeavor both 
theoretically and experimentally (for a comprehensive review see 
Ref.~\cite{jkg}).

Numerous complementary techniques have been developed in order to
detect relic WIMPs. Currently, the most promising 
method is probably direct detection through observation, in a 
low-background laboratory detector, of nuclear recoils due to
WIMP-nucleus elastic scattering~\cite{dirdet,annmod1}. 
The chance for a given WIMP to interact in the detector is 
very low and the energy released in case of interaction is expected 
to be tiny (in the keV range).
Nevertheless, this detection method has already had a few successes.
It has been exploited to exclude as the main component of the
dark Galactic halo WIMP candidates such as a fourth-generation
heavy neutrino and  the sneutrino~\cite{direxcl}.
Detectors have now reached the sensitivity to start probing 
the region of parameter space of interest if a neutralino
is the dark matter (see e.g. Ref.~\cite{dirnow}).
Recently, the {\sc Dama}~\cite{DAMA} and {\sc Cdms}~\cite{CDMS}
collaborations,
while probing roughly the same region of parameter space,
have presented apparently contradictory results, a possible WIMP
signal in the first case and a null result in the 
second. It is probably premature to derive any conclusion from 
these results, but, with further data and even more sensitive
detectors being developed, the next years promise to be very
exciting for the field.

To claim a positive detection, an experiment must be able to
discriminate the signal from backgrounds. In principle, 
the shape of the recoil spectrum can be used, since the recoil spectra
from WIMPs and background should generally differ.  However, the
shape of the recoil-energy spectrum for WIMP-nucleus scattering
cannot be predicted with enough precision to separate it from
the background, the spectrum of which is generally not
understood in detail.
A possible way out is to look for a slight annual 
modulation in the event rate
(see Refs.~\cite{annmod1,annmod2}; among more recent works see, 
e.g., Refs.~\cite{Brhlik,Belli}).
Such an effect is expected for the WIMP 
signal, but not for the background.
This is the signature exploited in the data analysis
by the {\sc Dama} collaboration to claim detection of WIMPs.
The underlying idea is quite simple. Like all other stars in the 
rotationally-supported disk, the Sun is moving around the Galactic
center on a roughly circular orbit, passing through the dark halo
which is believed on the other hand to be static and not
rotationally supported. The Earth, and detectors on it, contain
this velocity component plus an additional component due to the 
orbital motion around the Sun. The azimuthal velocity of the Sun
and the projection of the velocity of the Earth on the galactic 
plane are most closely aligned near June 2 and most 
anti-aligned six months later.
The WIMP-nucleus interaction rate in a detector depends
on the velocities of the incident WIMPs.  Hence, a yearly modulation
of the signal is expected.

In prior analyses of the modulation effect, the local dark-matter
velocity distribution was assumed to be a Maxwell-Boltzmann
distribution (which is, of course, isotropic), as would arise if 
the Galactic halo is isothermal.  Velocity distributions for
halos with some bulk rotation have also been considered~\cite{Belli}.
Although these velocity distributions are consistent with
current data on the Milky Way, there are other plausible,
consistent, and possibly even better-motivated alternatives.
For example, results from N-body simulations of hierarchical
clustering favor density profiles which are steeper
at large galactocentric distances than the 
$r^{-2}$ decline in the isothermal sphere and which
are cuspy in the Galactic center, rather than cored, and the velocity
distribution corresponding to a cuspy halo should differ from the
Maxwell-Boltzmann distribution that corresponds to an isothermal halo.

Moreover, it is plausible that the velocity distribution may be
anisotropic rather than isotropic as usually assumed.
In fact, most of the
visible populations in the Galactic halo show some degree of
anisotropy (e.g., the stars in the local neighborhood and
globular clusters).  Furthermore, the inefficiency of phase mixing
that results in a cuspy profile (rather than an isothermal
sphere) should leave some degree of anisotropy in the
velocity-dispersion tensor.
Some evidence for a global preference for predominantly
radial velocities is already seen in the 
simulations~\cite{VDB}, as well as in globular
clusters~\cite{OBGT}.   
Even if the global velocity distribution is isotropic, clumping in
velocity space, which may also arise if phase mixing is not perfectly
efficient during gravitational collapse, may yield a locally
anisotropic velocity dispersion.  

Prior work has shown that the direct-detection rate should not
depend sensitively on the details of the velocity distribution
\cite{Kink}.  However, this work considered only the {\it total} 
detection rate, integrated over all nuclear-recoil energies.  The
modulation signal in {\sc Dama} depends on details of the {\it
differential} energy distribution.  The purpose of this paper
will be to show that the amplitude of the modulation can thus
depend quite sensitively on the precise form of the velocity
distribution.  The inferred WIMP cross sections and masses could
thus be altered.

The outline of the paper is the following. In the next Section
we discuss a procedure to relate the velocity distribution 
to the Galactic density distribution. In Section~3 we review 
WIMP direct detection rates and the annual modulation effect.
The main results are given in Section~4. 
In Section~5 we summarize and make some concluding remarks.

\section{Dark-matter distribution functions}\label{sec:df}

We suppose that the dark-matter halo of the 
Milky Way is roughly spherical, and among the general family of 
profiles,
\begin{equation}
\rho_{\rm{dm}}(r) = \rho_0 \left(\frac{R_0}{r}\right)^{\gamma}
\left[\frac{1+(R_0/a)^{\alpha}}{1+(r/a)^{\alpha}}\right]^
{(\beta-\gamma)/\alpha},
\label{eq:dmprof}
\end{equation}
we focus on functional forms suggested by N-body simulations
(in the equation above $\rho_0$ and $R_0$ are respectively the 
local dark-matter density and the Sun galactocentric distance). 
We restrict ourselves mainly to the Navarro, Frenk, and White 
profile~\cite{NFW}, which has $(\alpha,\beta,\gamma)=(1,3,1)$ 
(hereafter the NFW profile). We will show also that the behavior
of the profile towards the Galactic center is not critical
in our analysis by considering the more cuspy 
Moore et al. profile~\cite{Moore}, 
$(\alpha,\beta,\gamma)=(1.5,3,1.5)$, and  the less singular profile
of Kravtsov et al.~\cite{Kravtsov},
$(\alpha,\beta,\gamma)=(2,3,0.4)$. The value of the scale 
radius $a$ which appears in Eq.~(\ref{eq:dmprof}) is determined 
in the N-body simulations as well, depending on the mass of the 
simulated halo.  We infer its approximate value for the NFW and 
Moore et al. profiles in case of an $\Omega_M = 0.3$ cosmology from 
Refs.~\cite{Moore,Bullock}. The approach we follow to fix the 
remaining unknown parameters, both in the dark-halo profile 
and in the functions that describe the luminous components
of the Milky Way, is to perform a combined best fit of 
available observational data, taking into account the kinematics
of local stars, the rotation curve of the Galaxy, the dynamics
of the satellites, and more (details are given in Ref.~\cite{Ullio1}).
Sample values for the subset of parameters relevant in the present
analysis are specified in Table~\ref{tab:rho} (mass 
decompositions for the Milky Way are highly
degenerate, so slightly different values are compatible as
well).  Thus, we have a family of spherically-symmetric radial
profiles that are all theoretically plausible and consistent
with all known observational constraints.

\TABLE[t]{%
  \begin{tabular}{|l|cccc|} \hline
  Parameter & $\rho_0$ & $a$ & $M_b$ & $M_d$ \\
  Units & GeV cm$^{-3}$ & kpc & 10$^{10}$ M$_{\odot}$ 
  & 10$^{10}$ M$_{\odot}$ \\ \hline
  NFW & 0.3 & 20 & 0.8 & 5.1 \\ 
  Moore et al. & 0.3 & 28 & 1.1 & 4.7 \\ 
  Kravtsov et al. & 0.6 & 10 & 1.3 & 3.6 \\ \hline
  \end{tabular}%
\caption{Best-fit values for some relevant parameters in a few 
mass decompositions for the Milky Way. The three dark-matter-halo 
profiles considered are specified in the text. 
$M_b$ and $M_d$ are respectively the total mass of the bulge and 
of the disk (stars + gas). We assumed that the Sun galactocentric
distance is 8~kpc.
}  \label{tab:rho}}

We will now find the velocity distributions that correspond to
these halo profiles.  The density distribution 
does not determine the velocity distribution uniquely.  To
sample the possibilities, we will therefore first find velocity
distributions that have isotropic velocity distributions, and
then find some distributions that have preferentially radial
velocities.

If we assume an isotropic velocity distribution, then 
there is a one-to-one correspondence between the spherically
symmetric density profile $\rho(r)$  and its distribution
function given by Eddington's formula~\cite{BT},
\begin{equation}
F(\eps) = \frac{1}{\sqrt{8} \pi^2}
\left[ \int_0^{\eps} \frac{d^2\rho}{d\Psi^2}
\frac{d\Psi}{\sqrt{\eps-\Psi}} + 
\frac{1}{\sqrt{\eps}} \left(\frac{d\rho}{d\Psi}\right)_{\Psi=0}
\right]
\label{eq:edd}
\end{equation}
where $\Psi(r) = - \Phi(r) + \Phi (r=\infty)$, with $\Phi$
the potential of the system, 
$\eps = - E + \Phi (r=\infty) = - E_{\rm{kin}} + \Psi(r)$, and
$E$ and $E_{\rm{kin}}$, respectively, the total and kinetic energy.
Eq. (\ref{eq:edd}) works for a single isolated self-gravitating
system.  However, the Milky Way has a complex structure
containing a bulge elongated into
a bar, a flattened disk, and maybe a triaxial dark halo. For the 
present purpose, however, it is sufficient to consider a toy model
in which all components are assumed to be spherical. Even the
awkward approximation of a ``spherical'' disk will have little
influence on our conclusions. 
In such a toy model, we can alter Eq.~(\ref{eq:edd}) to provide 
the dark-matter distribution function by replacing $\Psi$ and
$\rho$ (appropriate for an isolated system) by $\Psi_{\rm{tot}}$
and $\rho_{\rm{dm}}$, respectively, the gravitational potential due 
to all components and the dark matter density profile.
Actually, it is easier from the numerical 
point of view to implement Eq. (\ref{eq:edd}) by changing the
integration variable
from $\Psi_{\rm{tot}}$ to the radius of the spherical system $r$.  Then
Eq.~(\ref{eq:edd}), in case of the dark-matter halo 
distribution function, becomes,
\begin{eqnarray}
F_{\rm{dm}}(\eps) & = & \frac{1}{\sqrt{8} \pi^2}
\int_{\Psi_{\rm{tot}}^{-1}(\eps)}^{\infty} 
\frac{dr}{\sqrt{\eps-\Psi_{\rm{tot}}(r)}}
\left[\frac{d\rho_{\rm{dm}}}{dr} \frac{d^2\Psi_{\rm{tot}}}{dr^2}
\left(\frac{d\Psi_{\rm{tot}}}{dr}\right)^{-2} \right.
\nonumber \\
&& \left. -\frac{d^2\rho_{\rm{dm}}}{dr^2}
\left(\frac{d\Psi_{\rm{tot}}}{dr}\right)^{-1} \right]\; .
\label{eq:edd2}
\end{eqnarray}

If we relax the hypothesis of isotropy of the velocity dispersion
tensor, the most general distribution function corresponding to a 
spherical density profile is a function of $\eps$ and $L$, the 
magnitude of the angular-momentum vector. In such systems the 
velocity dispersion in the radial direction is different from
that in the azimuthal direction (which is equal to the one in 
the other tangential direction)~\cite{BT}. For a given radial density
profile, the distribution function is not unique. We 
investigate a special class of models, the Osipkov-Merritt
models~\cite{Osipkov,Merritt}, in which $F$ is a function of 
$\eps$ and $L$ only through the variable $\cal Q$:
\begin{equation}
{\cal Q} \equiv \eps - \frac{L^2}{2 r_a^2}\;.
\label{eq:q}
\end{equation}
Here $r_a$ is called the anisotropy radius, as in the Osipkov-Merritt
models the anisotropy parameter is~\cite{Merritt}:
\begin{equation}
\beta(r) \equiv 1 - \frac{\overline{v_{\phi}^2}}{\overline{v_{r}^2}}
= \frac{r^2}{r^2+r_a^2}\; .
\end{equation}
Therefore, $r_a$ is the radius within which the dispersion
velocity is nearly isotropic.
As already mentioned, in analogy with other observed populations,
we will entertain the possibility that for the dark-matter halo 
$\overline{v_{r}^2} > \overline{v_{\phi}^2}$ and thus that
$\beta > 0$.
The distribution function for this class of models is again
easy to derive. It is sufficient to replace in Eq.~(\ref{eq:edd2})
$\eps$ with $\cal Q$ and $\rho_{\rm{dm}}(r)$ with 
$\rho_{\rm{dm}}^{\cal Q}(r) = (1 + r^2/r_a^2)\, \rho_{\rm{dm}}(r)$.

\section{Direct-detection rates and annual-modulation effect}

The differential direct-detection rate for dark-matter WIMPs in
a given material (per unit detector mass) is~\cite{jkg},
\begin{equation}
\frac{dR}{dQ} = \frac{\rho_0}{M_{\chi}}
\int_{|\vec{v^{\prime}}| \geq v_{\rm{min}}} d^3\vec{v^{\prime}}\;
f(\vec{v^{\prime}}) \,|\vec{v^{\prime}}|\, \frac{d\sigma}{dQ} \;,
\label{eq:ddrate}
\end{equation}
where $Q$ is the energy deposited in the detector and $d\sigma/dQ$ 
is the differential cross section for WIMP elastic scattering with the 
target nucleus. We assumed here that WIMPs of mass $M_{\chi}$  
account for the local dark matter density $\rho_0$ and have a local
distribution in velocity space (in the rest frame of the detector)
$f = F_{\rm{dm}}(r=R_0)/\rho_0$. The lower limit of integration 
$v_{\rm{min}}$ is the minimum velocity required for a WIMP to deposit 
the energy $Q$.

Assuming that scalar interactions dominate (as is probably the case for
neutralino elastic scattering with Ge and NaI, the materials used
respectively by the {\sc Cdms} and {\sc Dama} experiments) and that 
the couplings with protons and neutrons are roughly the same, 
Eq.~(\ref{eq:ddrate}) can be rewritten as,
\begin{equation}
\frac{dR}{dQ} = \left(\frac{\rho_0\, 
\sigma_{p}^{\rm{scalar}}}{2}\right)
\left[ A_N^2 \frac{M_N}{{M_{\chi}}^3} 
\left(1+\frac{M_{\chi}}{M_p}\right)^2 {\cal F}^2(Q)\right]
\int_{|\vec{v^{\prime}}| \geq v_{\rm{min}}} d^3\vec{v^{\prime}}\;
\frac{f(\vec{v^{\prime}})}{|\vec{v^{\prime}}|} \; ,
\label{eq:ddrate2}
\end{equation}
where $\sigma_{p}^{\rm{scalar}}$ is the WIMP-proton cross section at 
zero momentum transfer, $A_N$ and $M_N$ are the detector nucleus 
atomic number and mass, while ${\cal F}(Q)$ is the nuclear form 
factor. 
In the equation above, the terms in the round bracket are energy and 
detector independent; we will not consider them in what follows. 
The terms in the square brackets depend on the nucleus chosen for the 
detector, as well as on the energy and WIMP mass.  When considering
annual modulation, they play a weighting effect for those 
detectors, like NaI, which are not monatomic (the generalization 
of Eq.~(\ref{eq:ddrate2})  to this case is straightforward).
The last term,
\begin{eqnarray}
T(Q,M_N,M_{\chi},t) &=& 
\int_{|\vec{v^{\prime}}| \geq v_{\rm{min}}} d^3\vec{v^{\prime}}
\; \frac{f(\vec{v^{\prime}})}{|\vec{v^{\prime}}|} =  
\int_{v_{\rm{min}}}^{\infty}  dv^{\prime} \;
v^{\prime} \int d\Omega^{\prime} \; f(v^{\prime},\Omega^{\prime}) 
\nonumber \\
&\equiv &
\int_{v_{\rm{min}}}^{\infty}  
dv^{\prime} \;g(v^{\prime}) \; ,
\label{eq:g}
\end{eqnarray}
depends on $Q$, $M_N$, and $M_{\chi}$ through 
$v_{\rm{min}}= [(Q M_N)/(2 M_r^2)]^{1/2}$, where $M_r$ is the 
WIMP-nucleus reduced mass. It is time dependent and gives rise 
to the annual-modulation effect.  This might not be clear at first 
sight, as we wrote implicitly the temporal dependence in the change of 
variables between the detector rest frame and the galactic frame.
In polar coordinates, the change of variable to the detector frame
(primed system in our notation) is simply:
$v_r = v_r^{\prime}$, $v_{\theta} = v_{\theta}^{\prime}$,
and $v_{\phi} = v_{\phi}^{\prime} + v_{\oplus}$. 
The azimuthal shift $v_{\oplus}$ varies 
during the year; in June it is roughly $v_{\oplus}= \Theta_0 + v_E$,
while in December it is $v_{\oplus}= \Theta_0 - v_E$, 
where $v_E \simeq 15 \,\rm{km}\, \rm{s}^{-1}$ is the projection of the
earth orbital velocity on the galactic plane, while $\Theta_0$ is 
the galactic circular velocity at the Sun's position. The latter is
given in terms of Oort's constants and the galactocentric
distance by:
\begin{equation}
\Theta_0 = (A - B) R_0 = (27.2 \pm 0.9) \, R_0 
\; \rm{km} \,\rm{s}^{-1} \,\rm{kpc}^{-1}\;,
\label{eq:theta}
\end{equation}
where the numerical value of $A - B$ comes from the determination
{}from Cepheid proper motions measured by the Hipparcos
satellite~\cite{FW}.

The amplitude of the annual modulation (keeping track of
whether the signal is greater in June or December) for a monatomic 
detector is then,
\begin{equation}
{\cal{A}}(Q,M_N,M_{\chi}) = 
\frac{T(Q,M_N,M_{\chi},{\rm June})-T(Q,M_N,M_{\chi},{\rm December})} 
{T(Q,M_N,M_{\chi},{\rm June})+T(Q,M_N,M_{\chi},{\rm December})}\;.
\end{equation}
As mentioned, the formula for NaI has instead a
weighting factor for each of the two nuclei. For a given detector and 
distribution function the value of $\cal{A}$ follows. To compute
$g$ as defined in Eq.~(\ref{eq:g}) the appropriate choice of 
integration variable are, in the isotropic case,
\begin{eqnarray}
g(v^{\prime}) & = & 2 \pi\,v^{\prime} 
\int_0^{\pi} d\alpha \; \sin\alpha\,
\frac{F_{\rm{dm}}(\eps)}{\rho_0}  \nonumber \\
\eps & = &\Psi(R_0) - \frac{1}{2} 
\left( {v^{\prime}}^2 + 2 \cos\alpha \, v^{\prime} v_{\oplus}
+ v_{\oplus}^2 \right) \;,
\end{eqnarray}
while in the anisotropic case,
\begin{eqnarray}
g(v^{\prime}) & = & 2\,v^{\prime} 
\int_0^{2 \pi} d\psi \int_0^{\pi} d\eta 
\; \sin\eta\,
\frac{F_{\rm{dm}}(\cal{Q})}{\rho_0}  \nonumber \\
\cal{Q} & = &\Psi(R_0) - \frac{1}{2} 
\left( {v^{\prime}}^2 + 2 \sin\psi\, \sin\eta \,v^{\prime} v_{\oplus}
+ v_{\oplus}^2 \right) \nonumber \\
&& - \frac{R_0^2}{2 r_a^2} \left( {v^{\prime}}^2\, \sin^2\eta 
+ 2 \sin\psi\, \sin\eta \,v^{\prime} v_{\oplus} + v_{\oplus}^2 \right)\;.
\end{eqnarray}

\section{Results}

\FIGURE[t]{
\epsfig{file=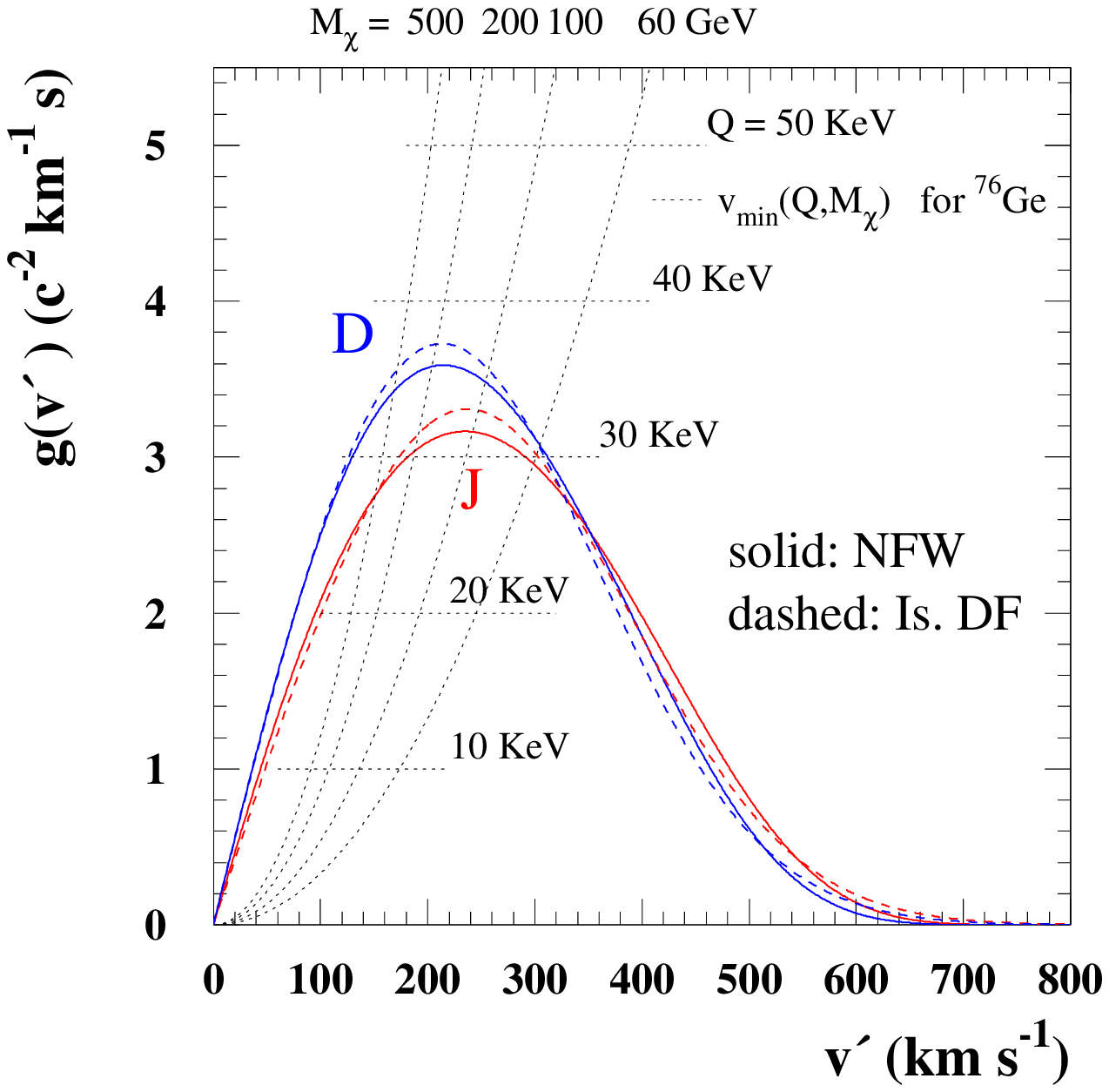,width=10.cm}
\caption{The integrand $g(v)$ for the direct-detection signal
for December (D) and January (J) for an NFW profile (solid) and
isothermal profile (dashed) both with an isotropic velocity
distribution.  The differential detection rate
for a given recoil energy $Q$ and WIMP mass $M_\chi$ is given by
the area under the curve to the right of the minimum velocity
$v_{\rm{min}}$.  This minimum velocity can be inferred in case of a
germanium detector from the
intersection of the dotted curves for a given $M_\chi$ and $Q$.
The annual-modulation amplitude is proportional to the
difference in the areas under the J and D curves for velocities
larger than $v_{\rm{min}}$.  The function $g(v)$ is the same for
other nuclei, but the dotted curves for $v_{\rm{min}}$ are
different.  Figs. \ref{fig:3}(a) and \ref{fig:4} show dotted
curves to determine $v_{\rm{min}}$ for Na and I, respectively.  A
galactocentric distance $R_0=8$ kpc was used.}
\label{fig:1}
}

\subsection{Isotropic Velocity Distributions}

We first consider distribution functions with isotropic velocity
dispersions. In Fig.~\ref{fig:1} we plot
with a solid line the function $g(v^{\prime})$ defined above
in case of a NFW profile, assuming the galactocentric distance
to be $R_0 = 8\; {\rm kpc}$ and $\Theta_0$ as derived from 
Eq.~(\ref{eq:theta}). 
There are two solid lines in the figure; the one which is higher 
at the peak refers to the function $g$ in December, while the 
second one 
is appropriate for June. As shown in the previous Section,
the amplitude ${\cal{A}}$ of the annual modulation is proportional to
the difference between June and December of the integral of $g$
above the value $v_{\rm{min}}$, which in turn depends on the energy
deposited in the 
detector and WIMP and nucleus masses. As a visual aid to identify
which are the relevant portions of the curves in each case,
we plot in the figure the value of $v_{\rm{min}}$ for a Germanium 
detector and a few values of $Q$ and $M_{\chi}$
(e.g. $v_{\rm{min}}(Q=30\;{\rm keV}, M_{\chi}= 60 \;{\rm GeV})$
is given by the abscissa of the point at the intersection between
the horizontal dotted line labeled $Q=30\;{\rm keV}$ and the
vertical dotted line labeled $M_{\chi}= 60 \;{\rm GeV}$).
Analogous plots for Na and I are given in 
Figs.~\ref{fig:3} and~\ref{fig:4}.

\FIGURE[t]{
\epsfig{file=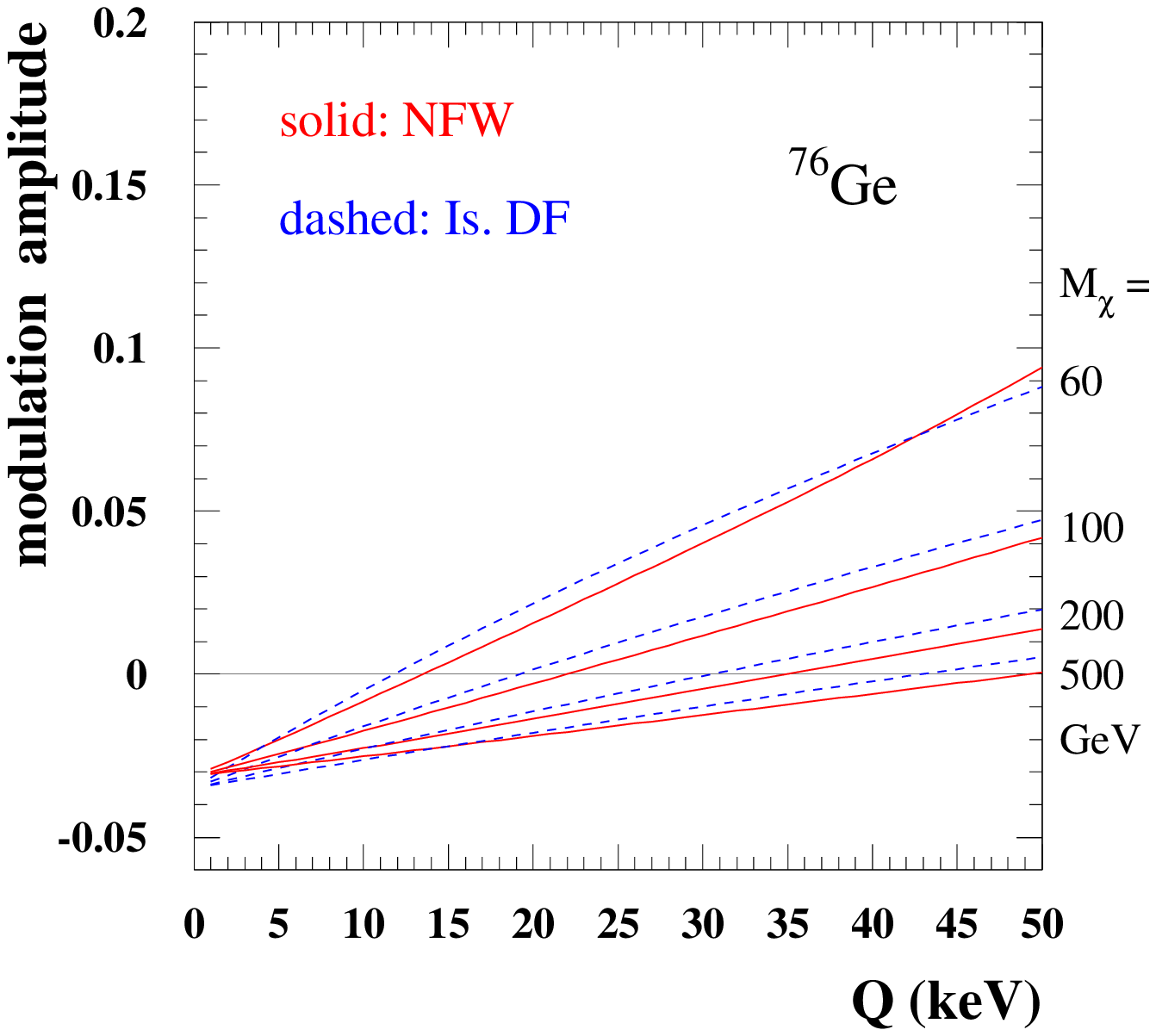,width=10.cm}
\caption{The annual-modulation amplitude for germanium as a function of
nuclear-recoil energy $Q$ for $R_0=8$ kpc for the NFW (solid) and
isothermal (dashed) profiles with isotropic velocity
distributions for several WIMP masses $M_\chi$. }
\label{fig:2}
}

In Fig.~\ref{fig:2} we plot the predicted annual-modulation
amplitude as a function of $Q$, for this NFW profile, for a
Germanium detector and
for four sample values for the WIMP mass. As known from previous
analyses, the modulation amplitude changes sign going to higher 
values of the deposited energy.
At least for low-mass WIMPs, the largest values of ${\cal{A}}$ 
correspond to the largest displayed value of $Q$.  Note however that 
at such large $Q$s the differential rate is almost negligible 
(being suppressed by the form factor ${\cal F}$). 

To compare with the case previous analyses focussed on, 
we display in Fig.~\ref{fig:1} the functions $g$ expected for 
an isothermal distribution function. The value for the velocity 
dispersion $\sigma$ is assumed accordingly to the naive 
(in the sense that it does not correspond to a self-consistent 
solution) prescription $\sigma = \sqrt{3/2} \;\Theta(R=\infty)$ and 
$\Theta(R=\infty) = \Theta_0$. We find a fairly good agreement 
with the NFW case and hence a consistency as well
in the values of the modulation amplitude in Fig.~\ref{fig:2}.

\FIGURE[t]{
\epsfig{file=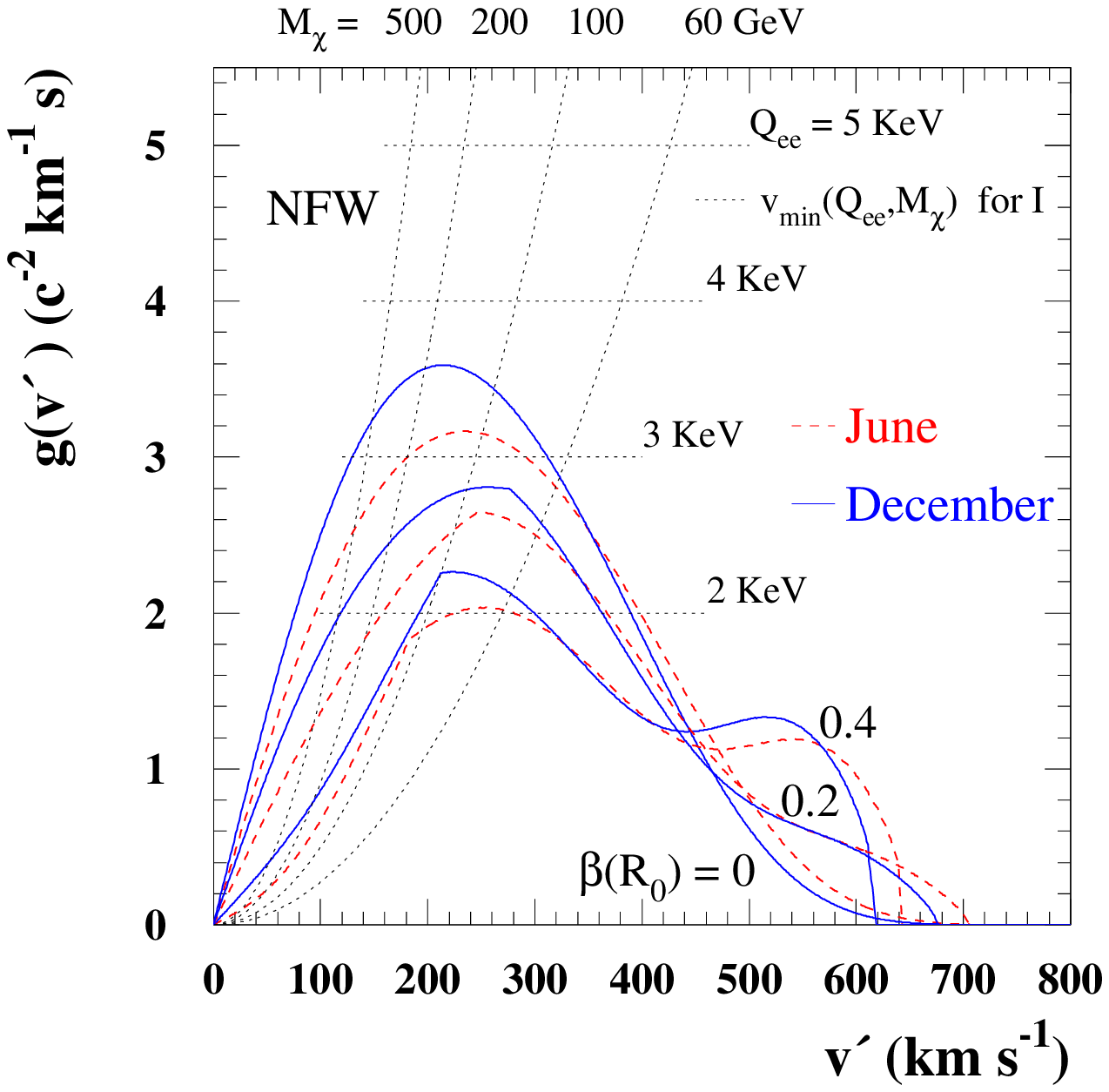,width=10.cm}
\caption{Like Fig. \ref{fig:1}, but for
anisotropic velocity distributions for the NFW radial profile.
Here, the solid curve is for December and the dashed curve is
for June, and curves are shown for several values ($\beta=0$,
0.2, and 0.4) of the anisotropy parameter.  The dotted curves
determine $v_{\rm{min}}$ for I.}
\label{fig:3}
}

We have checked that the effect we are trying to address does not 
depend sensitively on the steepness of the profile towards the 
Galactic center. The Moore et al. profile in Table~\ref{tab:rho}
gives curves for $g$ barely distinguishable from the NFW curves in  
Fig.~\ref{fig:1}. This is because these two profiles have 
similar amounts of dark matter inside $R_0$ and analogous ratios
of dark 
to luminous matter. The Kravtsov et al. profile in Table~\ref{tab:rho}
has best-fit values for the local halo density $\rho_0$ and for 
length scale $a$ respectively higher and lower than in the previous 
cases; the dark matter happens then to be appreciably more
concentrated toward the inner part of the Galaxy. 
In this case the velocity dispersion gets larger and hence
values for the modulation amplitudes are reduced. We go in the 
direction of a slightly larger velocity dispersion also by dropping 
the hypothesis of having a ``spherical disk''. 
We claim this in analogy to the isothermal case where it is
relatively easy to construct a self-consistent solution for a thin
disk and a flattened dark halo; this system has a velocity dispersion 
somewhat larger than in the purely spherical case.

\FIGURE[t]{
\epsfig{file=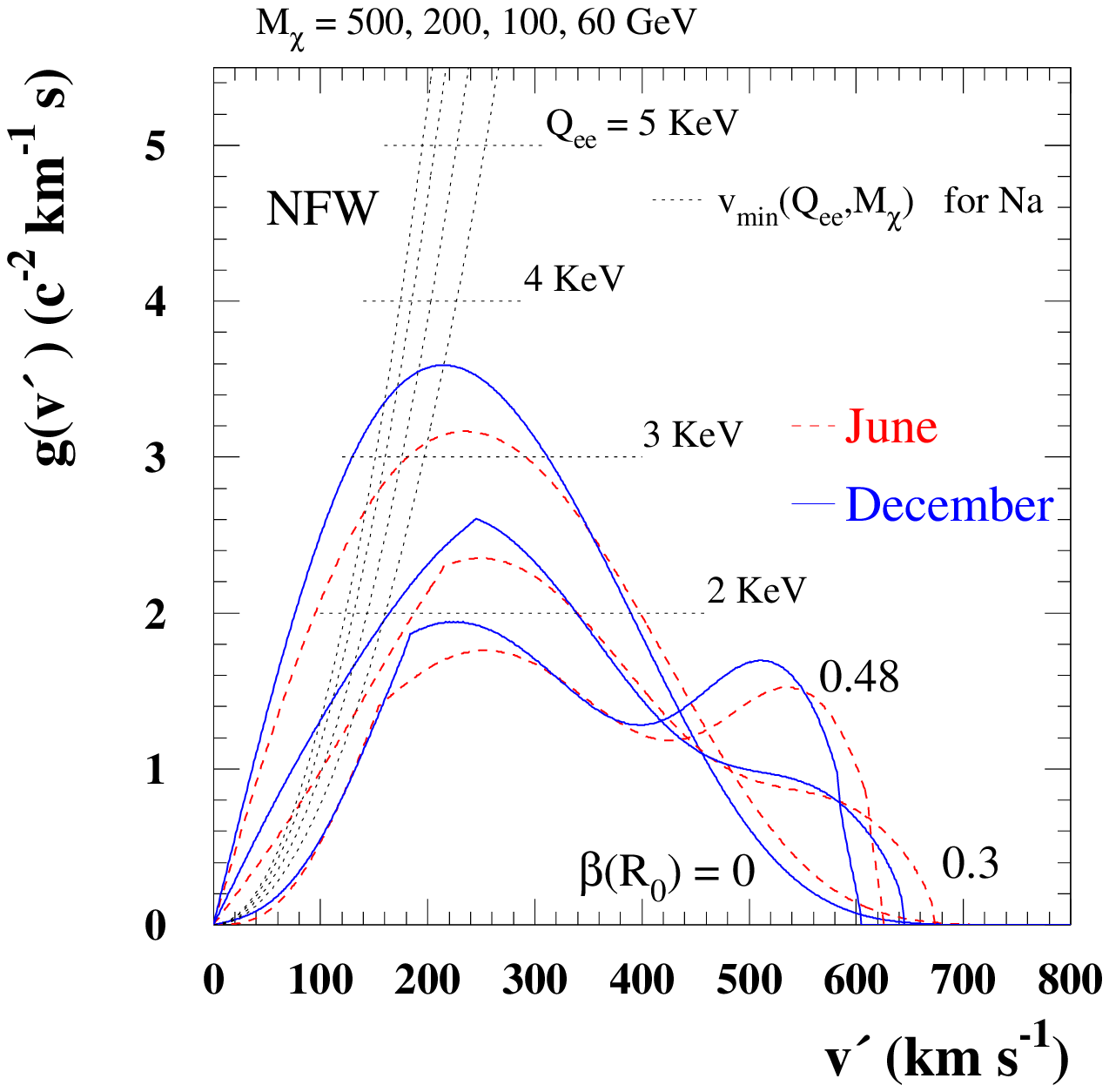,width=10.cm}
\caption{Like Fig. \ref{fig:3}, but for values ($\beta=0$,
0.3, and 0.48) of the anisotropy parameter. 
Again, the solid curve is for December and the dashed curve is
for June.  The dotted curves determine $v_{\rm{min}}$ for Na.}
\label{fig:4}
}

\subsection{Anisotropic Velocity Distributions}

We sketch now what happens in case of anisotropy in the velocity 
dispersion tensor. 
As mentioned in Section~\ref{sec:df}, different approaches 
are possible; we consider the Osipkov-Merritt models applied to
the NFW density profile introduced above. This will turn 
out to be sufficient to address the main qualitative effects. 
We suppose that the distribution function favors radial velocities. 
To illustrate the effects of anisotropy in the velocity
distribution, we consider values for 
the anisotropy parameter $\beta(R_0)$ in the range (0, 0.48).
The upper value is close to the value of $\beta(R_0)$ above
which, with our particular choice of potential and 
dark-matter-density profile, the Osipkov-Merritt scheme breaks down.

In Fig.~\ref{fig:3} we plot
the forms for the function $g$ corresponding to $\beta(R_0) = 0.2$
and 0.4, as well as $\beta(R_0) = 0$. In Fig.~\ref{fig:4} we plot
instead the $\beta(R_0) = 0.3$ and 0.48 cases.
It is evident that
detector rest-frame values of the WIMP kinetic energy are,
in anisotropic models, significantly redistributed (even though
the spherically symmetric density profile remains unaltered).
The enhancement at large $v^{\prime}$ is due to the fact that
in these models there is a higher probability to have a contribution
to the signal from particles on very elongated and nearly radial 
orbits (i.e. particles with $\cal{Q}$ close to zero; distribution 
functions analogous to the case considered here are given in 
Fig.~5 of Ref.~\cite{Widrow}). 
Obviously this implies that the recoil energy spectra changes 
to some extent.  However, without knowing the WIMP mass it may
be hard, in case of
a detection, to tell one spectrum induced by an anisotropic 
distribution from another due to an isotropic population for a
different WIMP mass.

\FIGURE[t]{
\epsfig{file=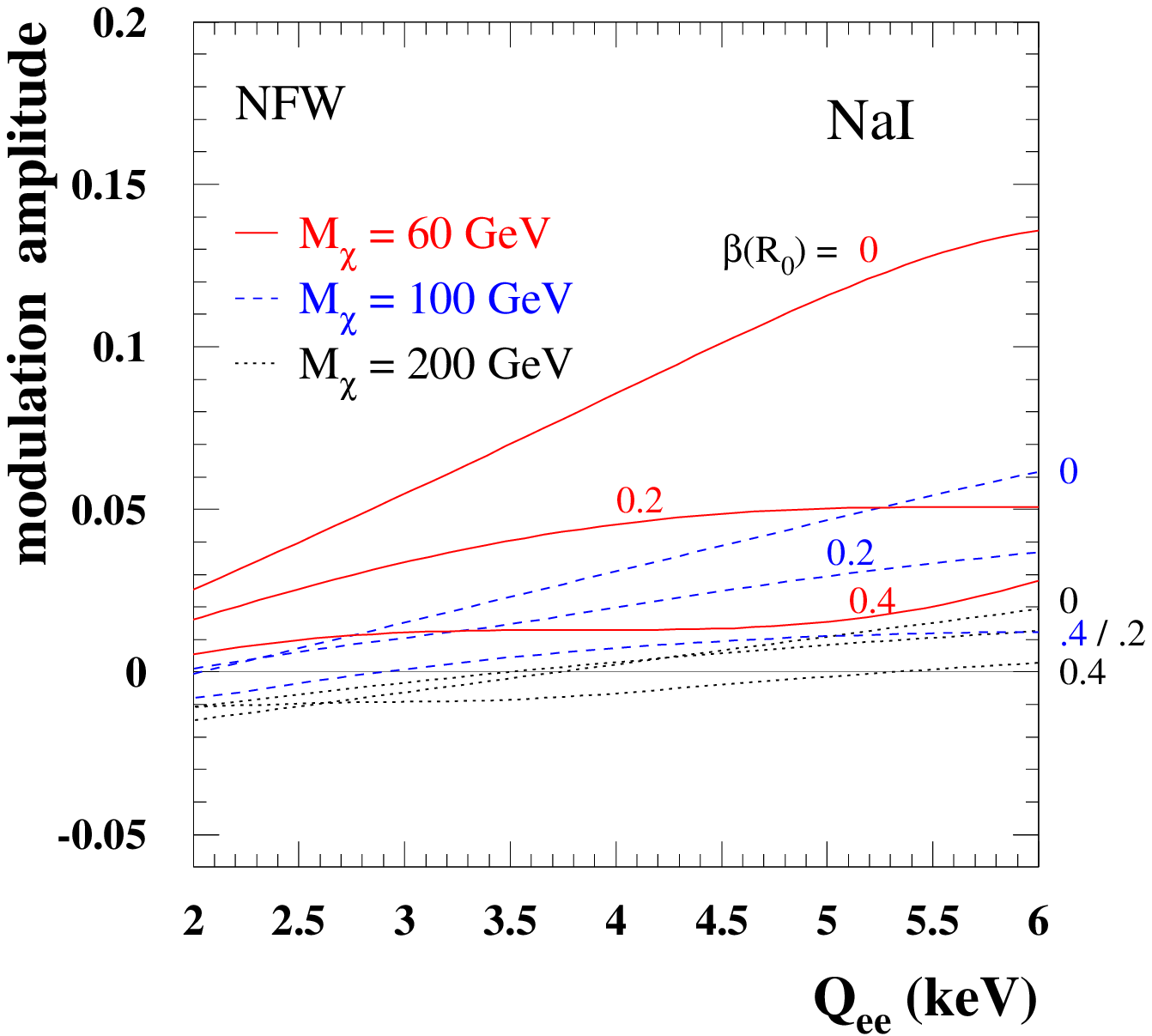,width=10.cm}
\caption{Like Fig. \ref{fig:2} but for Na instead of Ge and
for the NFW profile with anisotropic velocity distributions with
$\beta=0$, 0.2, and 0.4. Several different values of the
neutralino mass are considered. Note also that here we plot $Q_{ee}$,
the electron-equivalent energy that is measured by {\sc Dama}, 
rather than the nuclear recoil energy.
We see that the modulation amplitude may
depend on the anisotropy of the velocity distribution, and the
neutralino mass at which the modulation effect changes sign may
also depend on the velocity anisotropy.}
\label{fig:5}
}

Regarding instead the annual-modulation signature, the effect can 
be quite dramatic. 
In Fig~\ref{fig:5} we show the modulation amplitude for a NaI 
detector, plotting on the horizontal axis the electron equivalent 
energy $Q_{ee}$, rather than $Q$, in the range interesting for 
the {\sc Dama} experiment (to derive this plot we assumed  
quenching factors and Woods-Saxon form factors as suggested by 
the {\sc Dama} Collaboration~\cite{DAMA}). As can be seen,
for $\beta(R_0) = 0.4$, which corresponds to a local velocity 
ellipsoid for dark-matter particles that is still relatively close 
to spherical (the axis ratio is equal to 0.7), the modulation 
amplitudes in a NaI detector are severely damped. For intermediate 
neutralino masses and values of the recoil energy, the amplitude
even changes sign. The $\beta(R_0) = 0.2$ case lies between the
isotropic and 0.4 models.  Still, the influence on the annual 
modulation is still rather large. Results for a Ge detector 
and in case of $\beta(R_0) = 0.48$ and 0.3 are shown in 
Fig~\ref{fig:6} and are analogous to the NaI case.

\FIGURE[t]{
\epsfig{file=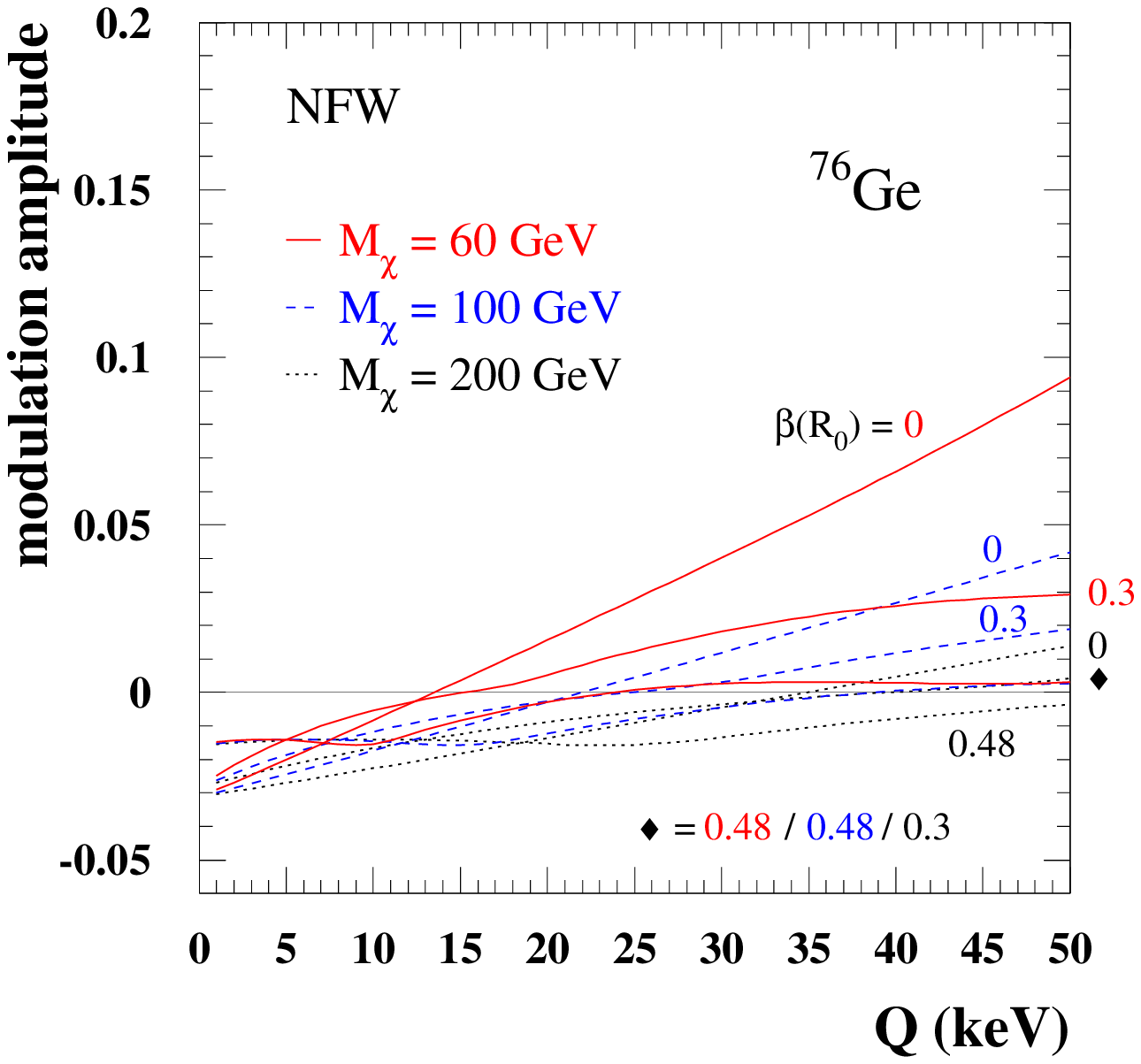,width=10.cm}
\caption{Like Fig. \ref{fig:4} but for Ge instead of Na. Here
NFW profiles with anisotropic velocity distributions with
$\beta=0$, 0.3, and 0.48 are considered.}
\label{fig:6}
}

\section{Conclusions}

We have investigated the possible effects on an
annual-modulation signal of additional structure in the WIMP
velocity distribution beyond the canonical Maxwell-Boltzmann
distribution.  To do so, we have considered isotropic
distribution functions that correspond to density profiles other 
than the isothermal profile that goes with a Maxwell-Boltzmann
velocity distribution, as well as some simple but plausible
anisotropic velocity distribution functions.

The measured local rotation curve of the Milky Way fixes
the velocity dispersion of any consistent dark-matter
phase-space distribution.  Thus, the {\it total} detection rate, 
integrated over all recoil energies, is expected to be
independent of the detailed form of the velocity distribution.
However, uncertainties in the velocity distribution can lead to
larger uncertainties in the predicted detection rate if a signal
is dominated primarily by events in a small recoil-energy bin,
as occurs, for example, in the {\sc Dama} annual-modulation
signal.  Moreover, the {\it sign}, as well as the magnitude, of
the annual modulation can be changed.  Thus, the constraints to
the WIMP mass that are inferred from the sign of the modulation
in {\sc Dama} may be loosened if we allow for some structure in
the phase-space distribution.

The anisotropic velocity distributions we used were chosen as
they provide simple deviations from a
Maxwell-Boltzmann distribution that illustrate
our point.  There are other possibilities for the phase-space
distribution that are more complicated, although perhaps better
motivated.  For example, the existence of an NFW
profile---rather than an isothermal profile---in numerical
simulations suggests that phase mixing via violent relaxation is
not fully efficient in gravitational collapse.  If so, then some
of the pre-collapse phase-space structure (recall that the
pre-collapse phase-space structure of cold dark matter is very
highly peaked around zero velocity) should be preserved.  It is
thus reasonable to expect some clumping in velocity space, even
if the halo is smooth in physical space.  Thus, for example, the 
{\it local} velocity distribution might be highly anisotropic
even if the velocity distribution averaged over a larger volume
of the Galaxy is isotropic.  Although numerical simulations will
be required to quantify this further, it is important to note
the possible implications for WIMP-detection rates with the
simple models we have considered.

\acknowledgments

We would like to thank George Lake for discussions.
This work was supported in part by NSF AST-0096023,
NASA NAG5-8506, and DoE DE-FG03-92-ER40701.


\end{document}